\documentclass{article}

     \PassOptionsToPackage{numbers, compress}{natbib}

     \usepackage[preprint]{neurips_2020}

\usepackage[utf8]{inputenc} 
\usepackage[T1]{fontenc}    
\usepackage[hidelinks]{hyperref}       
\usepackage{url}            
\usepackage{booktabs}       
\usepackage{amsfonts}       
\usepackage{nicefrac}       
\usepackage{microtype}      
\usepackage{natbib}
\usepackage{graphicx}
\usepackage{booktabs}  
\usepackage{array}     
\usepackage{multirow}  

\AtBeginDocument{%
  \providecommand\BibTeX{{%
    \normalfont B\kern-0.5em{\scshape i\kern-0.25em b}\kern-0.8em\TeX}}}
\title{Redefining Developer Assistance: Through Large Language Models in Software Ecosystem}

\author{
Somnath Banerjee$^1$, Avik Dutta$^1$, Sayan Layek$^1$, Amruit Sahoo$^1$\\ \textbf{Sam Conrad Joyce$^2$, Rima Hazra$^2$}\\
$^1$Indian Institute of Technology Kharagpur\\ 
$^2$Singapore University of Technology and Design\\
  \texttt{ \{som.iitkgpcse, sayanlayek2002\}@kgpian.iitkgp.ac.in}\\
 \texttt{ rima\_hazra@sutd.edu.sg} 
}

\begin{document}

\maketitle

\begin{center}
  \includegraphics[width=3.5cm, height=3.5cm]{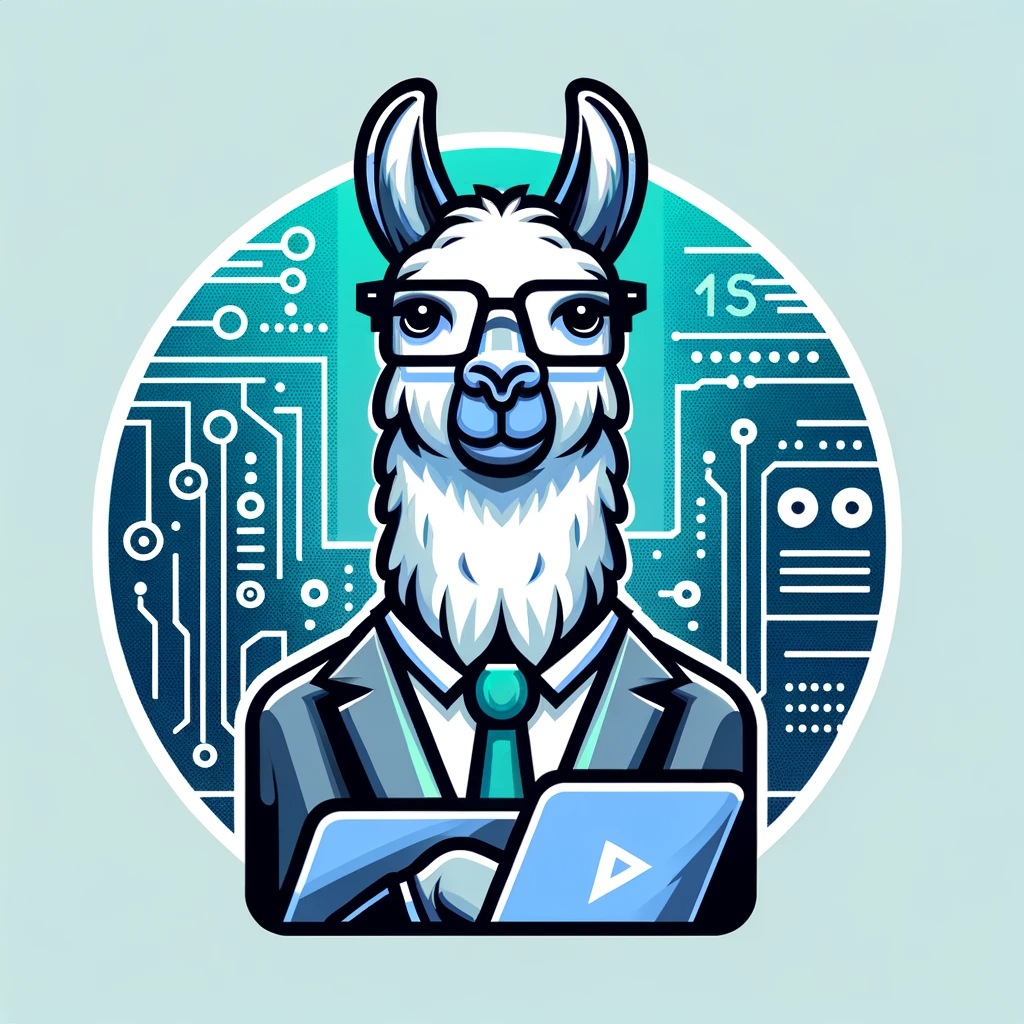}
  \end{center}
  \label{fig:teaser}
\begin{abstract}
In this paper, we delve into the advancement of domain-specific Large Language Models (LLMs) with a focus on their application in software development. We introduce DevAssistLlama, a model developed through instruction tuning, to assist developers in processing software-related natural language queries. This model, a variant of instruction tuned LLM, is particularly adept at handling intricate technical documentation, enhancing developer capability in software specific tasks. The creation of DevAssistLlama involved constructing an extensive instruction dataset from various software systems, enabling effective handling of Named Entity Recognition (NER), Relation Extraction (RE), and Link Prediction (LP). Our results demonstrate DevAssistLlama's superior capabilities in these tasks, in comparison with other models including ChatGPT. This research not only highlights the potential of specialized LLMs in software development also the pioneer LLM for this domain.
\end{abstract}

\maketitle

\section{Introduction}
The advent of Large Language Models (LLMs) has brought a revolutionary shift in various industries, with a significant impact in the field of software development. These models, ranging from General-Purpose Transformers like GPT by OpenAI~\cite{brown2020language} to specialized ones such as LLaMA~\cite{llamametaai} and Alpaca~\cite{alpaca}, have redefined human-computer interaction and automation in software engineering tasks. In the software industry, LLMs have been instrumental in automating tasks such as code generation~\cite{10.1145/3581641.3584037, zhang2023ecoassistant}, test case creation~\cite{schäfer2023empirical}, and program debugging~\cite{chen2023teaching}. Tools like OpenAI's Codex~\cite{chen2021evaluating} and GitHub Copilot~\cite{dakhel2023github} are exemplary in facilitating code-related tasks and enhancing developer productivity.\\
However, despite these advancements in the field, the software industry~\footnote{https://insights.sei.cmu.edu/blog/application-of-large-language-models-llms-in-software-engineering-overblown-hype-or-disruptive-change/} confronts substantial challenges beyond the realm of coding. While coding is an integral part of a developer's role, it occupies about 59.5\% of their work time~\footnote{https://www.software.com/reports/future-of-work}~\footnote{https://thenewstack.io/how-much-time-do-developers-spend-actually-writing-code/}, reflecting a notable focus on this aspect of software development. Surprisingly, a significant portion of a developer's time, nearly 40-45\%, involves tasks beyond coding. These include code maintenance~\footnote{https://thenewstack.io/how-much-time-do-developers-spend-actually-writing-code/}, testing, security issues~\footnote{https://www.techtarget.com/searchitoperations/feature/The-promises-and-risks-of-AI-in-software-development}, and other non-coding responsibilities that are essential for the holistic development of software. This division of time highlights the necessity for tools that cater not only to the coding aspect but also to the broader scope of software development. Tasks involving key concepts from Named Entity Recognition (NER), Relation Extraction (RE), Link Prediction (LP), Question Answering (QA), and Forum Answer Ranking (FAR) are pivotal in this broader scope~\cite{banerjee2024context,banerjee2024distalaner,hazra2023evaluating,hazra2024duplicate}. They play a critical role in interpreting complex technical documentation~\footnote{https://pub.aimind.so/interrogate-your-technical-documentation-using-free-and-paid-llms-f2a7664ff2bd}, managing software project requirements~\footnote{https://www.linkedin.com/pulse/requirement-engineering-using-llm-madhavan-vivekanandan/}, and facilitating effective communication within developer communities~\footnote{https://www.smehorizon.com/large-language-models-5-takeaways-for-smes/}.

Existing models fall short in addressing the unique requirements and complexities within the software domain considering facts. Their application in understanding technical specifications, extracting relevant information from software documentation, and assisting in the conceptualization of new software projects is not fully optimized. Moreover, in the realm of community engagement, particularly in online forums, the need for sophisticated models that can effectively manage and prioritize a wide range of tasks is increasingly evident. Current models have primarily focused on individual aspects of software development. However, the complexity and diversity of tasks in these forums demand a more integrated approach, combining all applicable tasks strengths.

Recognizing this gap, our research proposes the development of a comprehensive model DevAssistLlama~\footnote{Our model card?} designed to address these specific tasks in the software domain. This model aims to provide advanced unified solutions for NER, Relation Extraction, Link Prediction, Forum Answer Ranking, and Question Answering. The goal is to create a tool that not only enhances the efficiency of software developers but also provides nuanced and context-aware support in online software development communities. We explore the foundational technologies that empower these models and demonstrate how our proposed model is poised to reshape the paradigms of software development and community interaction, filling a critical gap in the current landscape.

\begin{figure}
\centering
\includegraphics[width=5.8cm]{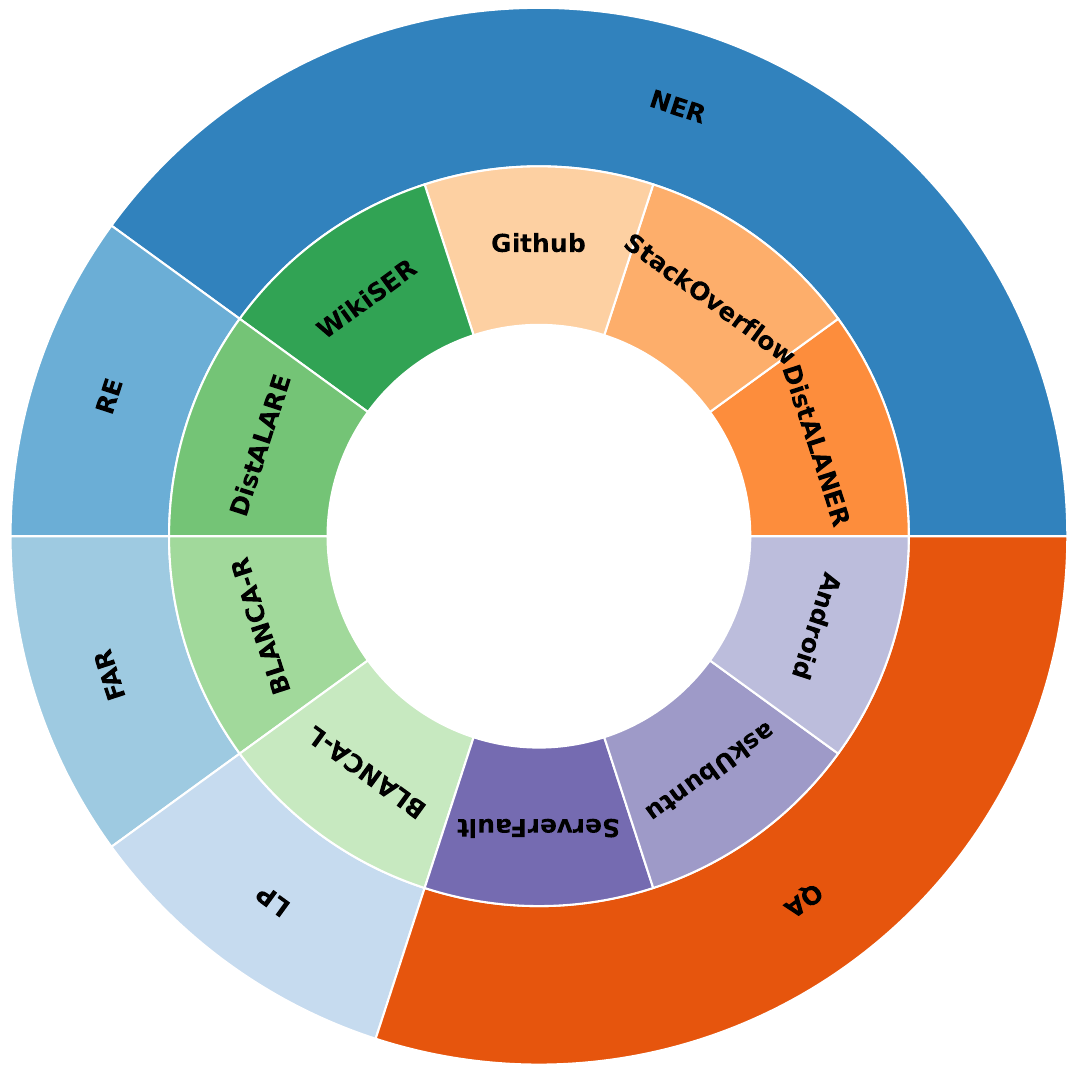}
\caption{Distribution of DevAssistLlama dataset}
\label{fig2}
\vspace{-0.2cm}
\end{figure}

\begin{table*}[!h]
\centering
\resizebox{.99\textwidth}{!}{
\begin{tabular}{>{\bfseries}lcccccc}
\toprule
\textbf{Models} & \multicolumn{4}{c}{\textbf{NER}} & \textbf{RE} & \textbf{LP} \\
\cmidrule(lr){2-5} \cmidrule(lr){6-6} \cmidrule(lr){7-7}
& DistALANER & StackOverflow & GitHub & WikiSER & DistALARE & BLANCA-L \\
\midrule
Alpaca         & 0.001 & 0.000 & 0.002 & 0.001 & 0.001 & 0.102 \\
Vicuna         & 0.028 & 0.000 & 0.003 & 0.000 & 0.008 & \textbf{0.899} \\
ChatGPT        & 0.350 & 0.420 & 0.300 & \textbf{0.450} & 0.400 & 0.850 \\
LlaMa 2 13B    & 0.010 & 0.001 & 0.001 & 0.002 & 0.150 & 0.812 \\
DevAssistLlama & \textbf{0.427} & \textbf{0.597} & \textbf{0.607} & 0.272 & \textbf{0.423} & 0.869 \\
\bottomrule
\end{tabular}
}
\caption{Quantitative Analysis of NLP Models on Named Entity Recognition, Relation Extraction, and Link Prediction.}
\end{table*}
\begin{table}[ht]
\centering
\begin{tabular}{>{\bfseries}lc}
\toprule
\textbf{Models} & \textbf{FAR} \\
\cmidrule(lr){2-2}
& BLANCA-R \\
\midrule
Alpaca     & 0.001 \\
Vicuna     & 0.165 \\
chatGPT    & 0.290 \\
LlaMa 2 13B & 0.297 \\
DevAssistLlama  & \textbf{0.312} \\
\bottomrule
\end{tabular}
\caption{FAR Comparison of Different Models}
\label{tab:far_data}
\end{table}

\begin{table}[ht]
\centering
\begin{tabular}{>{\bfseries}lccc}
\toprule
\textbf{Models} & \multicolumn{3}{c}{\textbf{QA}} \\
\cmidrule(lr){2-4}
& ServerFault & askUbuntu & Android \\
\midrule
Alpaca        & 0.42 & 0.48 & 0.44 \\
Vicuna        & 0.49 & 0.43 & 0.53 \\
chatGPT       & 0.76 & 0.72 & 0.64 \\
LlaMa 2 13B   & 0.76 & 0.76 & 0.71 \\
DevAssistLlama & \textbf{0.80} & \textbf{0.81} & \textbf{0.77} \\
\bottomrule
\end{tabular}
\caption{QA Performance of Different Models}
\label{tab:qa_performance}
\end{table}

\section{Dataset}
\begin{table}[ht]
\centering
\begin{tabular}{|p{0.5\linewidth}|p{0.4\linewidth}|}
\hline
\textbf{Dataset}             & \textbf{Number of Examples} \\ \hline
DistAlaner          & 450,000            \\
WikiSER             & 24,602             \\
StackOverflow       & 10,807             \\
Link Prediction     & 23,516             \\
Forum Answer Ranking & 330,151            \\
Android             & 16,523             \\
ServerFault         & 152,082            \\
AskUbuntu           & 134,655            \\ \hline
\end{tabular}
\caption{Dataset Statistics}
\label{tab:training_set}
\vspace{-0.2cm}
\end{table}

We mainly utilize our model for five state-of-art elementary tasks which are basic component of different high level tasks  -- named entity recognition, relation extraction, forum answer ranking, forum questions relateness, question-answering (See Figure~\ref{fig2}).  \\
\noindent \textbf{Named entity recognition (NER)}~\cite{anonymous_2023_8075578, Tabassum20acl, nguyen2023software}: NER is an elementary method for automated requirement extraction~\cite{10.1145/3427423.3427450}, code document analysis~\cite{chew2023llmassisted} and bug tracking and resolution. The task is to extract the entities of specific entity types given a software related textual information. We consider four software system related NER datasets. The ~\textsc{DistalaNER}~\cite{anonymous_2023_8075578} data consists of nine different software entity types. The textual information is the bug description. The entities are packages, operating system, organization, commands, error, file extension, peripheral, software component and computer architecture. Total number of instances present in the dataset is 170K. We used the ~\textsc{Stackoverflow} and ~\textsc{Github} data proposed in paper~\cite{Tabassum20acl}. In these datasets, the texts are the sentences of the questions. Both of them has around ~20 entity types where 8 code related entities and 12 natural language entities. The stackoverflow dataset consists of 15,372 sentences with 19,122 entities. The github data consists of 6,510 sentences with 10,963 entities. The Wikiser~\cite{nguyen2023software} dataset consists of 12 entity types. It contains the wikipedia textual information. This dataset consists of 1.7M sentences with 79K unique software entities.\\
\noindent \textbf{Relation extraction (RE)}~\cite{anonymous_2023_8075578}: In this dataset, textual information, along with two entities, serves as input, and the relation type acts as the output. It encompasses five relation types: dependency, conflict, affected version, cause and effect, and interaction/control. The textual information comprises bugs from the Ubuntu ecosystem. This dataset contains 500 instances. \\
\noindent \textbf{Forum answer ranking (FAR)}~\cite{abdelaziz2022blanca}: We use the forum answer ranking dataset from the Blanca benchmark proposed in paper~\cite{abdelaziz2022blanca}. The task is to predict the best answers relative to other answers provided for a given question in a forum. The dataset consists of 500K questions, each having at least 3 answers, with an average of ~4.9 answers per question. \\
\noindent \textbf{Forum link prediction (LP)}~\cite{abdelaziz2022blanca}: The task involves predicting whether two questions are related or not. The dataset consists of 23,516 pairs of posts, with 11,758 related pairs and 11,758 unrelated pairs.\\
\noindent \textbf{Question answering}~\footnote{ \url{https://serverfault.com/}}~\footnote{\url{https://askubuntu.com/}}~\footnote{\url{https://android.stackexchange.com/}}: 
The task requires the model to generate answers for software-related questions. We compile datasets from three community question answering systems: Android CQA with 21,560 pairs, AskUbuntu with $\sim$175,000 pairs, and ServerFault with $\sim$210,500 pairs, all from early 2023.

\section{Developer assistance through instruction tuning}
In this work, our main objective is to build a model which would assist developers to address natural language software queries in various stages. We utilize the instruction tuning method on pretrained LLMs to achieve this objective.
Instruction tuning is one of the fine tuning methods for aligning the llm to specific tasks or specific domain. For each pair of instruction and desired response \((x, y)\) in the instruction dataset, we update the model parameters \(\theta\) to reduce the difference between the model's output \(y^{'}\) and desired output \(y\). Further, this process iteratively update \(\theta\) to improve the model's performance in following instructions and producing the desired outputs. In our work, this methodology involves the refinement of pre-existing llama2-13b. 
There are various existing works on instruction tuning where the main focus is to solve specific tasks~\cite{zhou2023universalner, Lu_Zhao_Mac}. Whereas we investigate how these pretrained LLMs as well as instruction tuned LLM can be employed as a component in the pipeline of various software system tasks. In the subsequent sections, we describe the construction of instruction dataset, instruction tuning process and evaluation tasks.

\begin{table}[h!]
\centering
\small
\begin{tabular}{|>{\raggedright\arraybackslash}p{0.9\columnwidth}|}
\hline
\textbf{System Message} \\
{[}INST{]}You work with software developers from software industries. Your task is to \{\textbf{Task}\}.{[\textbackslash}INST{]} \\
\hline
\textbf{Task: \{Named Entity Recognition\}} \\
\textbf{Instructions:} \\
Assume that you are a named entity recognizer. You will be given text, entity type as inputs and you have to extract possible entities of given entity type from the text. \\
\hline
\textbf{Task: \{Link prediction\}} \\
\textbf{Instructions:} \\
Assume that you are a software question relevance checker. You will be given two questions as inputs and you have to identify whether two questions are relevant or not. Use 0 for non-relevant and 1 for relevant. \\
\hline
\textbf{Task: \{Community Question Answering\}} \\
\textbf{Instructions:} \\
Assume that you are a helpful software community question answering expert. You will be given a question as inputs and you have to provide the answer. Your answer should be very crisp and clear. If you don't know the answer, do not try to make up the answer. \\
\hline
\textbf{Task: \{Forum Answer Ranking\}} \\
\textbf{Instructions:} \\
Assume that you are an answer ranker in a community question answering portal. Given the question, rank the following answers based on their relevance from most relevant (1) to least relevant (0). If an answer is completely irrelevant, it should be ranked last. Strictly follow the Output Format: "OPTION [answer number]: Rank [ranking number]". DO NOT include any other additional text. \\
\hline
\end{tabular}

\caption{System Tasks and Instruction Prompts}
\label{instSet}
\end{table}

\subsection{Instruction dataset construction}
For instruction tuning, we build an instruction dataset from various software system datasets. 
This dataset is meticulously curated to encompass a broad spectrum of instructional formats, contexts and ensuring comprehensive exposure to various types of related tasks. 
We build instruction dataset from different datasets of major tasks in software systems such as named entity recognition, forum answer generation, forum answer ranking and forum related questions prediction. Firstly, we merge training datasets present for all these tasks. We provide training dataset statistics in Table ~\ref{tab:training_set}. Secondly, we create instruction for each of the tasks and prepare the instruction set. We provide prompts of each task in Table ~\ref{instSet}. 
Further, we shuffle each datasets and considered 5000 instances from each dataset. Our final training set contains 45000 instances with five different type of instructions.

\subsection{Instruction tuning}
After preparing the instruction dataset, we implement instruction tuning on the Large Language Model (LLM) to specifically adapt it for various tasks commonly handled by software developers. This fine-tuning involves inputting both the definitions of tasks and relevant textual information into the LLM. Such an approach ensures that the model becomes finely attuned to the particular nuances and specific requirements of domain-centric tasks.

\subsection{Negative sampling}
Our methodology crucially includes negative sampling in the instruction dataset to enhance the model's ability to differentiate between relevant and irrelevant or non-existent data. In Named Entity Recognition (NER) tasks, for example, we introduce scenarios where text falsely associates with a non-existent entity type, expecting a null response from the model. This trains the model to recognize both the presence and absence of entities, reducing false positives. For instance, if a text mentions software tools but no programming languages, and the entity type is 'programming language', the model should produce an empty output. Similarly, for question-answering tasks, we include questions outside the software system domain, like historical or culinary queries, to teach the model to filter out irrelevant inquiries. Negative sampling is vital for enhancing the model's precision and reliability, equipping it to handle scenarios where the absence of relevant information is crucial. This approach ensures accurate responses and avoids misleading information when the model lacks necessary context.

\section{Experimental setup}
\subsection{Model setup}
In our study, we employ a model setup inspired by~\cite{vicuna2023}. The cornerstone of our experimentation is Llama 2, fine-tuned to enhance its efficacy in our designated tasks. This approach is in line with cutting-edge methodologies in natural language processing. For evaluation, we select up to 1,000 instances from three distinct random splits, ensuring a diverse and representative test dataset. This strategy aims to emulate real-world scenarios and offers a comprehensive evaluation of the model's capabilities.
\subsection{Compared models}
We evaluate the performance of several prominent Large language models. ChatGPT  (gpt-3.5-turbo-0301) is renowned for its conversational abilities and versatility in various language tasks, initially based on the GPT-2 model from~\cite{radford_language_2019}. Vicuna, credited to~\cite{vicuna2023}, stands out for its efficiency, being fine-tuned on a LLaMA base model using conversational data. Similarly, Alpaca, developed by~\cite{alpaca}, uses the LLaMA architecture but is specifically tuned for domain-specific dialogues. Our focus, Llama 2, is fine-tuned following~\cite{touvron2023llama} guidelines, showcasing its capabilities in complex language processing scenarios and offering insights into the evolving landscape of vaious NLP tasks.
\subsection{LoRA approach for fine-tuning}
Low-Rank Adapters (LoRA) represent a pivotal advancement in the field of natural language processing, particularly in the fine-tuning of large language models. First introduced by~\cite{hu2021lora}, LoRA allows for the efficient and effective modification of pre-trained models, significantly reducing the computational overhead typically associated with training large-scale neural networks. In our implementation, we integrate LoRA adapters directly into the attention mechanisms and train both the language model (LM) heads and the adapters concurrently, diverging from traditional methods that prioritize the initial training of embeddings alone, as seen in models like the Chinese LLaMA~\cite{cui2023efficient}. Our infrastructure utilizes the robust capabilities of six NVIDIA A100 GPUs, each with 46GB of memory, enabling us to train our model, DevAssistLlama, over three epochs spanning a total of 63 hours. We showcased important hyperparameters in Table~\ref{tab:hyperparameters}. 

\section{Results}
Our evaluation of various models in the NER, RE, and LP tasks, measured using the macro F1 score, offers significant insights, as detailed in Table 1. In the NER tasks across DistALANER, StackOverflow, GitHub, and WikiSER datasets, DevAssistLlama leads with scores of 0.427, 0.597, 0.607, and 0.272, respectively. ChatGPT also shows impressive performance with scores of 0.350, 0.420, 0.300, and 0.450 in the same tasks, indicating its strong capabilities in named entity recognition.

For the RE task on DistALARE, DevAssistLlama tops with a score of 0.423, closely followed by ChatGPT with a score of 0.400. In the LP task on BLANCA-L, DevAssistLlama again leads with a score of 0.869, with ChatGPT showing a commendable score of 0.850.

\noindent \textit{FAR Performance:} Table 2 illustrates the FAR performance, using the Mean Reciprocal Rank (MRR) as the evaluation metric. Here, DevAssistLlama achieves the highest MRR score of 0.312, with LlaMa 2 13B at 0.297. ChatGPT's performance is also noteworthy, with an MRR score of 0.290, showcasing its efficiency in fault analysis and recovery.

\noindent \textit{QA Performance:} In QA tasks covering ServerFault, askUbuntu, and Android domains (Table 3), DevAssistLlama outshines others with scores of 0.80, 0.81, and 0.77, respectively. ChatGPT, however, delivers the highest performance in this category, with impressive scores of 0.76, 0.72, and 0.64, indicating its strong suitability for QA tasks.

\noindent \textit{Overall Assessment:} This comprehensive evaluation across diverse NLP tasks using macro F1 score, MRR for FAR, and FactSUMM for QA tasks, highlights the exceptional capabilities of DevAssistLlama and ChatGPT. DevAssistLlama consistently tops in NER, RE, and LP tasks, while ChatGPT excels notably in QA tasks, affirming its proficiency in complex language processing challenges.

\subsection{Evaluation metrics}
To evaluate our model, we employ a range of metrics to assess performance across different tasks. For Named Entity Recognition (NER), Relation Extraction (RE), and forum link prediction tasks, we utilize the F1 score. This metric is preferred in machine learning for its ability to assess predictive skill by focusing on class-wise performance. The F1 score combines precision and recall, using their harmonic mean, and is particularly useful in situations where datasets are class-imbalanced~\footnote{https://tinyurl.com/adkbk2yk}.

For tasks like forum answer ranking, we apply the Mean Reciprocal Rank (MRR). MRR is a measure that determines the mean of the reciprocal ranks of the actual relevant item in the ranking list of items provided in response to queries~\footnote{https://en.wikipedia.org/wiki/Mean\_reciprocal\_rank}. For question answering datasets, we use factsumm metric which measure the facts overlap between actual answer and generated answer.

\begin{table}[h]
\centering
\begin{tabular}{|l|}
\hline
\textbf{Important hyperparameters of DevAssistLlama} \\ \hline
Learning Rate: 2e-4| 
Max sequence length: 4096| \\
LoRA Alpha: 16| Training Precision: FP16| \\
Epochs: 3| Optim: paged\_adamw\_32bit| \\
fp16=True| bf16=True| Warmup ratio=0.03| \\
LR scheduler=Cosine| Trainer= SFTT| \\
Per device training batch=8 \\ \hline
\end{tabular}
\caption{Summary of important hyperparameters}
\label{tab:hyperparameters}
\vspace{-0.4cm}
\end{table}

\section{Limitation and future work}
The DevAssistLlama, a fine-tuned version of the Llama 2 model specifically for the software domain, exhibits certain limitations and areas for future development.

\noindent\textbf{Contextual Understanding:} While DevAssistLlama may excel in general language processing, its understanding of complex software terminology and concepts could be limited. This could affect its ability to accurately interpret and respond to highly technical queries.

\noindent\textbf{Code Generation and Debugging:} The model might struggle with generating error-free code or providing accurate debugging solutions, as these tasks require a deep understanding of programming logic and syntax.

\noindent\textbf{Data Bias and Quality:} The model's performance heavily depends on the quality and diversity of the training data. Any biases or limitations in the dataset can lead to suboptimal or skewed outputs.

\noindent\textbf{Real-time Adaptation:} DevAssistLlama might face challenges in adapting to the rapidly evolving software technologies and terminologies, leading to outdated or irrelevant responses.

To address these limitations we can explore certain steps.
\noindent\textbf{Enhanced Technical Training:} Integrating more comprehensive and diverse software-specific datasets could improve the model's understanding of complex technical concepts and terminologies.
\noindent\textbf{Collaboration with Developers:} Working closely with software developers to fine-tune the model could lead to more accurate and contextually relevant outputs.
\noindent\textbf{Real-Time Learning Mechanisms:} Implementing mechanisms for real-time learning and adaptation could help the model stay updated with the latest software trends and practices.
\noindent\textbf{Robust Code Generation and Debugging:} Focusing on enhancing the model's capabilities in code generation and debugging could make it a more valuable tool for software development.

\noindent

\bibliographystyle{plainnat}
\bibliography{references}







\end{document}